\documentclass[11pt,leqno]{article}

\usepackage{mathptmx}
\usepackage{amsmath,amsthm,amssymb}
\usepackage{enumerate} 
\usepackage{graphicx}

\theoremstyle{plain}
\newtheorem*{thm}{Theorem}

\makeatletter

\makeatletter

\makeatletter

\makeatletter

\title{\Large\bf On a law of large numbers for insurance risks}
\author{Yumiharu Nakano\footnote{
This study is partially supported by JSPS KAKENHI Grant Number 26800079.
}
\\[1em]
        \small{Graduate School of Innovation Management} \\
        \small{Tokyo Institute of Technology} \\
        \small{2-12-1 W9-117 Ookayama 152-8552, Tokyo, Japan}
}

\date{\today}

\begin{document}

\maketitle

\begin{abstract}
This note presents a kind of the strong law of large numbers for an insurance risk caused by 
a single catastrophic event rather than by an accumulation of independent and identically 
distributed risks. We derive this result by a large diversification effect resulting from optimal 
allocation of the risk to many reinsurers or investors.  

\begin{flushleft}
\end{flushleft}
\end{abstract}


It is well-known that the expected value premium principle is 
justified by the law of large numbers. 
Namely, if an insurance company has a collection of risks $W_1,W_2,\ldots$, which is 
independent and identically distributed, then by the strong law of large numbers  
in probability theory, we have 
\begin{equation*}
 \lim_{n\to\infty}\frac{1}{n}\sum_{i=1}^nW_i = \mathbb{E}[W_1]
\end{equation*}
with probability one. Thus the averaged risk is approximately amount to the capital 
given by the expectation of the individual risk. 
This means in general that by collecting independent risks and by adding the 
expected value premium we can decrease the uncertainty of the risk to zero eventually. 

The objective of this note is to ask how about the case where the risk is caused by 
a single catastrophic event rather than by an accumulation of a large number of independent 
events. In this case, of course, we cannot apply the strong law of large numbers in probability 
theory. Instead, we approach to this problem by considering a diversification of 
the single risk to a large number of reinsurers or investors. 

To formulate the problem mathematically, 
suppose that an insurance company with initial capital $w_0$ has a risk with 
loss random variable $X$. Here we assume that $X$ is bounded. 
Then the resulting position of the company at a terminal time 
becomes $w_0-X$. Suppose moreover that this company would like to diversify this risk partly to 
another $n$ reinsurers or investors, 
each of which has an initial capital $w_i$, $i=1,\ldots,n$, 
as a reinsurance or a securitization. 
Let us denote by $X_0$ the risk that the originator covers, and 
by $X_i$ the one allocated to the $i$-th reinsurer or investor, for $i=1,\ldots,n$. 
Each participant is willing to accept this risk if the resulting expected utility is 
greater than or equal to that of the initial capital.  
The allocation problem is thus defined as 
\begin{equation}
\label{eq:1}
\begin{split}
 &\max\mathbb{E}[u_0(w_0-X_0)],  \\
 &\text{s.t.} \;\; \mathbb{E}[u_i(w_i -X_i)]\ge u_i(w_i), \quad i=1,\ldots,n, 
\end{split}
\end{equation}
over all bounded random variables $X_i$ such that $X_0+\cdots + X_n=X$. 
Here, $u_0$ denotes the utility function of the insured, and 
$u_i$ the one of the $i$-th reinsurers, $i=1,\ldots,n$. 
Note that in order the problem (\ref{eq:1}) to be feasible, the allocated random variables 
$X_i$, $i=1,\ldots,n$, necessarily take negative values, 
which include benefits of the transaction for the participants. 

The problem (\ref{eq:1}) is categorized as an optimal design of insurance policies or 
more generally as a risk sharing problem.  
See Borch \cite{bor:1960}, \cite{bor:1962}, Raviv \cite{rav:1979}, 
Br{\"u}hlmann \cite{buh:1970}, Gerber \cite{ger:1979}, 
and the references therein. 
It should be remarked that to our best knowledge, 
the limiting analysis with respect to the number of agents is 
not studied in the literature except Fukuda et.al~\cite{fuk-etal:2009} 
where some dynamic insurance pricing is examined.  

Let us denote by $\hat{X}_i\equiv \hat{X}_i^{(n)}$, $i=0,1,\ldots,n$, a solution of (\ref{eq:1}). 
Then we will claim that $\hat{X}_0^{(n)}$ converges to $\mathbb{E}[X]$ as 
$n\to\infty$ with probability one. 
This implies that the optimal covering of the risk for the originator is approximately 
the expectation of the risk, so the expected value premium principle is justified provided that  
the risk is sold to a large number of investors. 

To solve (\ref{eq:1}), first we assume that each player has an exponential utility function: 
\begin{equation*}
 u_i(x) = \frac{1}{a_i}(1-e^{-a_i x}), \quad x\in\mathbb{R}, \;\; i=0,\ldots,n, 
\end{equation*}
with $a_i>0$, $i=0,\ldots,n$. 
Then, for a given $\lambda_i>0$, $i=1,\ldots,n$, we consider 
\begin{equation}
\label{eq:2}
 \max\sum_{i=0}^n\lambda_i\mathbb{E}[u_i(w_i-X_i)]
\end{equation}
over all bounded random variable $X_i$, $i=0,\ldots,n$ such that 
$X=X_0+\cdots +X_n$, where we have set $\lambda_0=1$. 
Suppose that $\{\hat{X}_i\}_{i=0}^n$ is a solution to the problem (\ref{eq:2}) 
for some $\{\lambda_i\}_{i=1}^n$ such that 
\begin{equation}
\label{eq:3}
 \mathbb{E}[u_i(w_i-\hat{X}_i)] = u_i(w_i), \quad i=1,\ldots,n. 
\end{equation}
Then for $\{X_i\}$ satisfying $X=\sum_{i=0}^nX_i$ and the constraints in (\ref{eq:1}), 
we have 
\begin{align*}
 \mathbb{E}[u_0(w_0-X_0)] &\le \mathbb{E}[u_0(w_0-X_0)]+\sum_{i=1}^n\lambda_i\left(
 \mathbb{E}[u_i(w_i-X_i)]-u_i(w_i)]\right) \\
 &=\sum_{i=0}^n\lambda_i\mathbb{E}[u_i(w_i-X_i)] -\sum_{i=1}^n\lambda_iu_i(w_i)
 \le \sum_{i=0}^n\lambda_i\mathbb{E}[u_i(w_i-\hat{X}_i)] -\sum_{i=1}^n\lambda_iu_i(w_i) \\
 &= \mathbb{E}[u_0(w_0-\hat{X}_0)].  
\end{align*} 
Hence $\{\hat{X}_i\}_{i=0}^n$ is a solution to (\ref{eq:1}). 
Therefore, our task is to solve (\ref{eq:2}) for any $\{\lambda\}_{i=1}^n$ and then 
choose $\{\lambda\}_{i=1}^n$ to satisfy (\ref{eq:3}). 

Now, by Borch's risk exchange theorem (see, e.g., Gerber and Pafumi \cite{ger-paf:1998} 
for a review with notation similar to ours), 
the optimal $\{\hat{X}_i\}_{i=0}^n$ exists and necessarily satisfies 
\begin{equation*}
 \lambda_i u_i^{\prime}(w_i-\hat{X}_i) = \lambda_j u_j^{\prime}(w_j-\hat{X}_j), 
 \quad i,j=0,1,\ldots,n, 
\end{equation*}
almost surely. 
Thus the random variable 
\begin{equation*}
 \Lambda = \lambda_i e^{-a_iw_i}e^{a_i\hat{X}_i}
\end{equation*}
does not depend on $i=0,\ldots,n$, and so
\begin{equation}
\label{eq:4}
 \hat{X}_i = \frac{1}{a_i}(\log\Lambda -\log\lambda_i) + w_i, \quad i=0,\ldots,n. 
\end{equation}
Summing up the both side over $i$ and then using $X=\sum_i X_i$, we have 
\begin{equation*}
 X = \frac{1}{a}\log\Lambda - \sum_{i=1}^n\frac{\log\lambda_i}{a_i} + w, 
\end{equation*} 
where $a$ is defined by $1/a = \sum_{i=0}^n (1/a_i)$ and $w=\sum_{i=0}^nw_i$. 
From this, 
\begin{equation*}
 \log\Lambda = a(X-w) + a\sum_{i=1}^n\frac{\log\lambda_i}{a_i}. 
\end{equation*}
Substituting this into (\ref{eq:4}), we obtain 
\begin{equation*}
 \hat{X}_i = \frac{a}{a_i}(X-w) - \frac{\log \lambda_i}{a_i} + w_i + \frac{a}{a_i}\sum_{j=1}^n 
  \frac{\log\lambda_j}{a_j}, \quad i=0,\ldots,n. 
\end{equation*}

Now consider the case where $a_1=\cdots =a_n$ and 
$w_1=\cdots =w_n$ for any $n$. Then we can expect $\hat{X}_1=\cdots=\hat{X}_n$ 
and so it is natural to assume $\lambda_1=\cdots=\lambda_n$. 
In this case, the equalities just above become 
\begin{gather}
 \hat{X}_0 = \frac{a}{a_0}(X-w) + w_0 +\frac{na}{a_0a_1}\log\lambda_1, \label{eq:5} \\
 \hat{X}_1 = \frac{a}{a_1}(X-w) + w_1 - \frac{\log\lambda_1}{a_1} 
   + \frac{na}{a_1^2}\log\lambda_1. \label{eq:6} 
\end{gather}
Further, the condition (\ref{eq:3}) implies   
\begin{equation*}
 \mathbb{E}\left[\frac{1}{a_i}(1-e^{-a_i(w_i-\hat{X}_i}))\right] 
 = \frac{1}{a_i}(1-e^{-a_iw_i}),  \quad i=1,\ldots,n,  
\end{equation*}
and by (\ref{eq:6}) we have 
\begin{equation*}
 \mathbb{E}[e^{a_1\hat{X}_1}] = \mathbb{E}[e^{-aX}]e^{-aw+a_1w_1}\lambda_1^{-(1-na/a_1)}. 
\end{equation*}
So, (\ref{eq:3}) for $i=1$ is equivalent to 
\begin{equation}
\label{eq:7}
 -\left(1-\frac{na}{a_1}\right)\log\lambda_1 + \log\mathbb{E}[e^{-aX}] 
 -aw +a_1w_1 = 0. 
\end{equation} 
This equality determines $\lambda_1$ that we look for. 
(\ref{eq:6}) and (\ref{eq:7}) yield 
\begin{align*}
 \hat{X}_1&= \frac{a}{a_1}(X-w) + w_1+\frac{1}{a_1}\left(a_1w_1- \left(1-\frac{na}{a_1}\right)
  \log\lambda_1 \right) \\
  & = \frac{a}{a_1}(X-w) + \frac{1}{a_1}\left(-\log\mathbb{E}[e^{aX}] +aw\right) 
  = \frac{a}{a_1}X - \frac{1}{a_1}\log\mathbb{E}[e^{aX}].   
\end{align*}
Further, 
\begin{equation*}
 \hat{X}_0 = \frac{a}{a_0}(X-w) + w_0 
   - \frac{na}{a_0a_1}\frac{1}{1-\frac{na}{a_1}}\left(\log\mathbb{E}[e^{aX}] -aw + a_1w_1\right). 
\end{equation*}
Here, $a_1/(na)=(na_0+a_1)/(na_0)$. Thus, 
$1-na/a_1=a_1/(na_0+a_1)$. From this we see 
\begin{equation*}
 \frac{na}{a_0a_1}\frac{1}{1-\frac{na}{a_1}} 
 = \frac{na}{a_0a_1}\left(1+\frac{na_0}{a_1}\right) 
 = \frac{na}{a_1}\cdot\frac{1}{a} = \frac{1}{a} - \frac{1}{a_0}. 
\end{equation*}
Therefore, (\ref{eq:5}) and (\ref{eq:7}) yield 
\begin{equation*}
 \hat{X}_0 = \frac{a}{a_0}X + \left(\frac{1}{a}-\frac{1}{a_0}\right)\log\mathbb{E}[e^{aX}] 
 - \frac{aw}{a_0} + w_0 -\frac{naw}{a_1} + nw_1 
 = \frac{a}{a_0}X + \left(\frac{1}{a}-\frac{1}{a_0}\right)\log\mathbb{E}[e^{aX}]. 
\end{equation*}
Now, since $a\to 0$ as $n\to\infty$ and 
\begin{equation*}
 \lim_{s\to 0}\frac{1}{s}\log\mathbb{E}[e^{sX}] = \mathbb{E}[X], 
\end{equation*}
we finally arrive at the following result: 
\begin{thm}
Under the assumptions and the notation above, 
the optimally diversified risks $\hat{X}_0\equiv\hat{X}_0^{(n)}$ 
and $\hat{X}_1\equiv\hat{X}_1^{(n)}$ satisfy 
\begin{equation*}
 \lim_{n\to\infty}\hat{X}_0^{(n)} = \mathbb{E}[X], \quad 
 \lim_{n\to\infty}\hat{X}_1^{(n)} = 0, 
\end{equation*}
almost surely. 
\end{thm}
The theorem means that under the assumptions that each agent has an exponential utility and 
the investors for the diversification are homogeneous, 
by diversifying the risk over many reinsurers or investors and by adding the 
expected value premium we can decrease the uncertainty of the risk to zero eventually.
Moreover, the participants for the diversification lose nothing in the limit. 

\bibliographystyle{plain}
\bibliography{mybib}

\end{document}